\documentclass[
aip,
pop,
 amsmath,amssymb,
 reprint,%
]{revtex4-1}
\usepackage{hyperref}
\usepackage{graphicx}
\usepackage{dcolumn}
\usepackage{bm}
\usepackage{multirow}
\usepackage[utf8]{inputenc}
\usepackage[T1]{fontenc}
\usepackage{mathptmx}
\usepackage{etoolbox}

\makeatletter
\def\@email#1#2{%
 \endgroup
 \patchcmd{\titleblock@produce}
  {\frontmatter@RRAPformat}
  {\frontmatter@RRAPformat{\produce@RRAP{*#1\href{mailto:#2}{#2}}}\frontmatter@RRAPformat}
  {}{}
}%
\makeatother
\begin{document}

\preprint{AIP/123-QED}

\title{Intensity and Dimensionality-Dependent Dynamics of Laser-Proton Acceleration in 1D, 2D, and 3D Particle-in-Cell Simulations}
\author{Lillian A. Daneshmand}
 \affiliation{Department of Physics and Astronomy, University of Iowa, Iowa City, IA, USA}\affiliation{Physics Department, Marietta College, Marietta, OH, USA}
 
\author{Madeline Aszalos}%
 \affiliation{Physics Department, Marietta College, Marietta, OH, USA}

\author{Scott Feister}%
\affiliation{California State University Channel Islands, Camarillo, California 93012, USA}

\author{Joseph R. Smith}
  \email{JosephRSmith@protonmail.com}
 \affiliation{Physics Department, Marietta College, Marietta, OH, USA}

\date{\today}

\begin{abstract}
Due to the high computational cost of 3D particle-in-cell (PIC) simulations, lower-dimensional (2D or 1D) simulations are frequently used in their place. Our work shows that when modeling high-intensity laser ion acceleration, simulation dimensionality interfaces with laser intensity in the dynamics of ion acceleration at every step of the process, from laser absorption through particle acceleration. We expand on previous studies by comparing the behavior of 1D and 2D simulations (of different polarization) with 3D PIC simulations at high resolutions across five orders of magnitude of laser intensity, enabling us to study multiple regimes of laser-proton acceleration. We find that key output metrics such as maximum proton energy depend on a complex interplay of both simulation dimensionality and laser intensity regime. Differences between simulation predictions generally increase for higher laser intensity regimes, making 3D simulations especially important for quantitative predictions of next-generation laser experiments.

\end{abstract}

\maketitle

\section{\label{sec:intro}Introduction}
Particle-In-Cell (PIC) simulations enable researchers to better understand intense laser-plasma interactions\cite{ziegler2024laser} and to probe unexplored regimes, paving the wave for future advances in laser technology~(e.g.,~\cite{zhang2015effect,jirka2017qed}). However, lower dimensional simulations\cite{Wagner_2016,higginson2018near} are often used in place of 3D simulations, which may require thousands of CPU cores\cite{fonseca2013exploiting,ngirmang2020evidence,bird2021vpic}, or hundreds to thousands of GPUs\cite{hilz2018isolated,myers2021porting,bird2021vpic,garten2023temporal,ziegler2024laser}. Trends in 1D and 2D laser-proton acceleration simulations often persist in 3D\cite{ferri2019enhanced,fedeli2020enhanced,smith2020optimizing,goodman2022optimisation}, although they generally overestimate the maximum proton energy compared to 3D simulations\cite{sgattoni2012laser,liu2013three,d2013optimization,babaei2017rise, xiao2018multidimensional}. Additionally, the choice of polarization in 2D simulations impacts laser absorption and the acceleration process~\cite{liu2013three,stark2017effects,stark_2018,gu2021multi}. We build upon previous work by exploring five orders of magnitude of laser intensity and analyzing how differences in laser-plasma conversion efficiency and energy transfer rates determine variations in key output metrics in the simulations. We show the behavior and outcomes of 1D, 2D, and 3D PIC simulations of high-intensity laser ion acceleration are influenced by a complex interaction between the simulation dimensionality and laser intensity regime.

\begin{figure}
    \centering
    \includegraphics{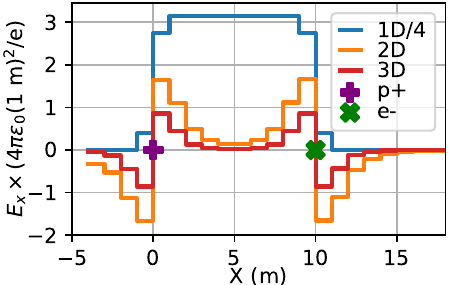}
    \caption{Electric field due to a proton at the origin and an electron at $X$=10~m in 1D, 2D, and 3D PIC simulations. The electric field drops off more slowly in 2D and not at all in 1D. The 1D electric field is divided by 4 to display results on a similar scale. [Associated dataset available at \url{https://doi.org/10.5281/zenodo.13338477}.] (Ref.~\cite{dataset_zenodo}).     }
    \label{fig:field_drop}
\end{figure}

There are a number of fundamental differences to consider when comparing lower dimensional PIC simulations to higher dimensional simulations and real-world experiments. Figure~\ref{fig:field_drop} demonstrates the electric field drop off in PIC simulations with different dimensionality\footnote{This demonstration uses the default triangle shape function of the EPOCH PIC code and a cell size of 1~m in each dimension.}. In 3D, the long-range drop off of electric fields for isolated charges behave as expected with an $E\propto 1/r^2$ dependence. In 2D\footnote{We only consider Cartesian spatial grids in one, two, or three physical spatial dimensions in this comparison.}, the particles behave like an infinite line charge with  $E\propto 1/r$ dependence, and in 1D the particles behave like infinite sheets of charge where the electric fields do not drop at large distances ($E\propto r^0 = 1$)~\cite{birdsall2018plasma, touati2022kinetic}.

Clearly, these unphysically high electric fields at far distances change the rates of energy transfer between particle species in PIC simulations as explored in this paper. Additionally, Gaussian laser pulses come to a focus more gradually in lower dimensional simulations~\cite{Ngirmang_2016} and the field and particle energy reported in lower dimensional simulations must be scaled to directly compare to 3D\cite{Smith_2021_Compare}. Further discussion and formulae used to scale our results are included in Appendix~\ref{app:focus_energy}.

Section~\ref{sec:sim_setup} presents the simulation parameters used in the work. Next we explore differences in the laser ion acceleration process over 5 orders of magnitude in laser intensity in Sec.~\ref{sec:physical_processes}. Consequences of dimensionality and intensity on the key output metric of maximum proton energy are explored in Sec.~\ref{sec:output_metrics} and we conclude in Sec.~\ref{sec:conclusion}.

\section{Simulation Parameters}\label{sec:sim_setup}

\begin{figure}
    \centering
    \includegraphics[width=\linewidth]{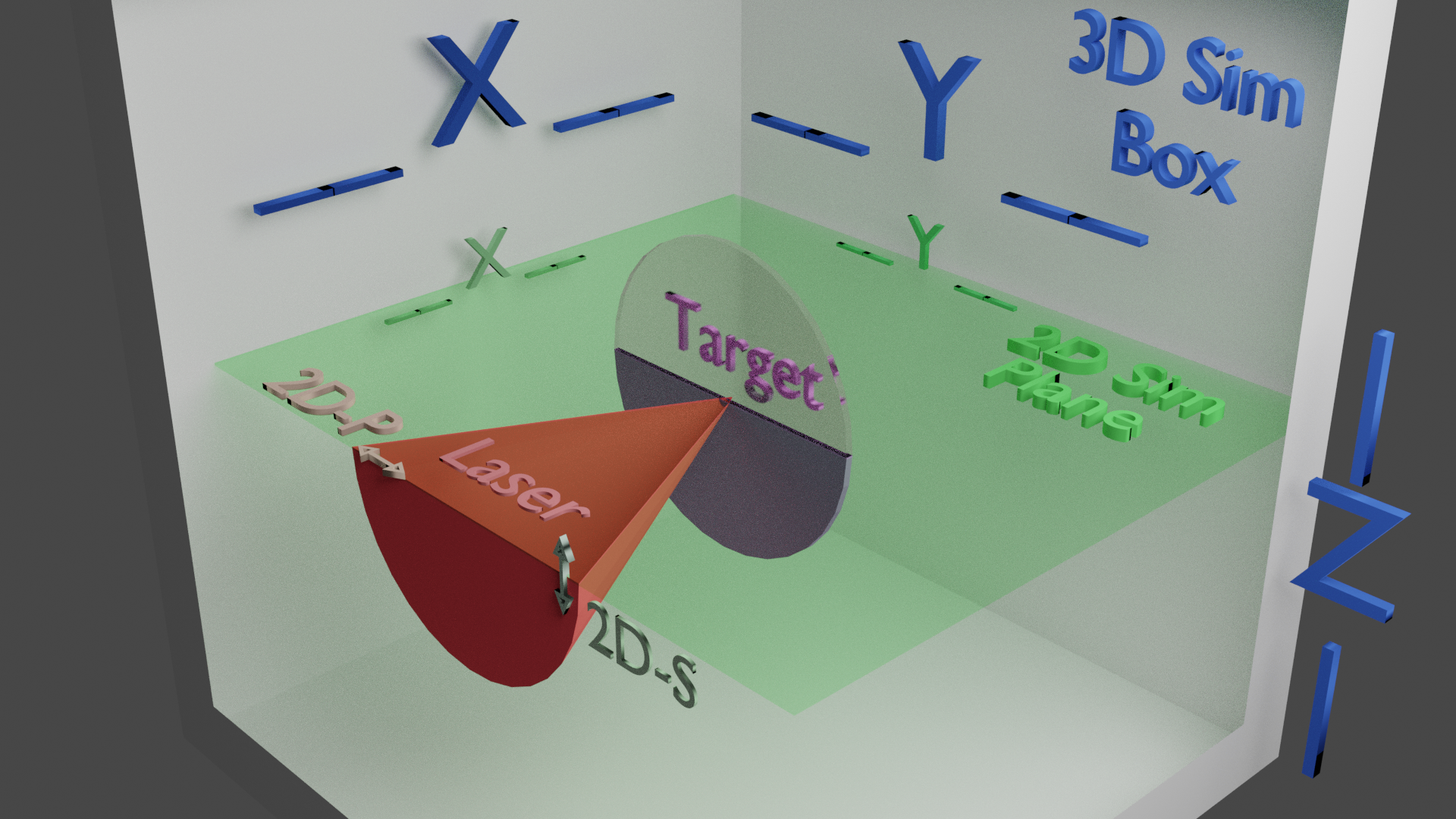}
    \caption{Schematic showing the shared geometry for all simulations. The 3D simulation box (or 2D simulation plane/1D simulation line) extends from -42 $\mu m$ to +42 $\mu$m along all relevant axes ($X/Y/Z$ for 3D, $X/Y$ for 2D, and $X$ for 1D). The coordinate origin is coincident with both the laser's focal point and the target's center (in all dimensions). The laser propagates in the +$X$ direction. For 3D simulations and 2D-S simulations, the laser is polarized in the +$Z$ direction. For 2D-P simulations, the laser is polarized in the +$Y$ direction.}
    \label{fig:SetUpSchematic}
\end{figure}

The PIC code EPOCH \cite{Arber:2015hc} is used to run simulations in 1D, 2D, and 3D.  The general simulation setup is based on the one from \citet{Smith_2021_Compare}, where three different PIC codes show strong agreement. The `prototype' input deck used for all simulations is provided in the supplementary material, and simulation inputs, output data, and analysis scripts are available through Zenodo\cite{dataset_zenodo}. As shown in Fig.~\ref{fig:SetUpSchematic}, the simulation box size is 42 $\mu$m per side in each dimension, and the cell size was 20 nm in each dimension. The 800 nm wavelength laser pulse propagates along the $X$-axis with a spatial profile of a Gaussian and a sine-squared temporal profile with a 30 fs FWHM. The beam waist radius for these simulations is $w_0=$ 2 $\mu$m.

These simulations include a 1 $\mu$m thick target composed of ionized hydrogen, the radius of the target (in 2D/3D) was set to 10 $\mu$m. The target was given uniform proton and electron densities of 8.5 $\times$ 10$^{21}$~cm$^{-3}$  and 50 macroparticles per cell per species were used. An initial target temperature of 1~eV was used for all simulations to limit the initial particle energy in lower intensity simulations. The Debye length is not resolved given the initial temperature, but significant numerical heating was not observed in these simulations.  The 20~nm cell size used in these simulations is large enough to allow for the plasma skin depth, in this case 58 nm, to be resolved\cite{Smith_2021_Compare}.

The maximum simulation time for all simulations was 1000 fs, although some 3D simulations were stopped prior to this. The default timestep of 0.95 times the Courant–Friedrichs–Lewy (CFL) limit\cite{CFL_Paper,CFL_Paper_English} was used. This means that the timestep was slightly smaller for higher dimensional simulations since, for a grid with side length $\Delta x$ in each dimension, the CFL limit is $CFL = \Delta x/(c\sqrt{N})$, where $N=1,2,3$ is the dimensionality and $c$ is the speed of light \cite{Arber:2015hc}.

When simulating a linearly polarized pulse in 2D, the laser can be polarized in a physical dimension ($Y$ for our simulation), or in the virtual dimension ($Z$)\footnote{The laser could also be polarized at some angle between these two.}. At normal incidence, there is no real-world difference between these choices, but there are non-trivial differences in simulation predictions\cite{stark2017effects} that are further explored in this work. The plane of incidence is ill-defined for a normally incident laser pulse, although it is common to use 2D-P and 2D-S to describe in-plane polarization and out of plane polarization respectively for simplicity \cite{stark2017effects}.

\subsection{Regimes of Laser Ion Acceleration}
We explore how differences between simulations of different dimensionality depend on the laser peak intensity, by running nine simulations in each of the four configurations from intensities of 10$^{17}$~W~cm$^{-2}$ up to 10$^{21}$~W~cm$^{-2}$. To understand the regime of the laser interaction it is useful to refer to the normalized laser amplitude
\begin{equation}
    a_0 \equiv \frac{e E_0}{m_{e}\omega c} \approx 0.85 \sqrt{\frac{I\lambda^2}{10^{18}~W~cm^{-2}}},
    \label{a0}
\end{equation}
where $E_0$ is the peak electric field amplitude, $m_e$ is the electron mass, and $\omega = 2 \pi c/\lambda$ is the laser angular frequency \cite{macchi2013superintense}. When $a_0$ exceeds unity (about 2$\times$10$^{18}$~W~cm$^{-2}$ for our simulations), the laser pulse is considered `relativistic.'

The classical critical density of a target is given by $n_{\rm c} = {\varepsilon_0 m_e \omega^2}/{e^2} \approx 1.7 \times 10^{21}~\rm{cm}^{-3}\times( 0.8~\mu m / \lambda)^2$. Our target is approximately five times the classical critical density. For relativistic lasers of sufficient intensity, a classically opaque target can become transparent due to an effect known as Relativistic Induced Transparency (RIT). The relativistic correction to the critical density is $n_{\rm c^\prime} =\gamma n_{\rm c}$, where the relativity factor of the plasma, which can be estimated as $\gamma = \sqrt{1 + {a_0^2}/{2}}$.\cite{chen2016introduction,macchi2013superintense,king2023perspectives} For our laser and target conditions, the threshold for RIT effects occurs at  $a_0 \approx 6.8$, or a laser intensity of approximately 10$^{20}$~W~cm$^{-2}$. We note that this does not mean that our targets will become fully transparent at this intensity, rather this provides a rough lower bound for a new regime of laser ion acceleration. A more complex treatment of relativistic transparency, such as for targets thinner than the laser wavelength may be found in Refs.~\cite{Vshivkov_1998,macchi2013superintense}.

\subsection{Data Reduction}
Due to the large number of high resolution 3D simulations, we utilize several data reduction techniques. At a macro-scale the total energy in the simulation, including energy stored in both the particles and the fields was recorded every 5 fs. We used `subsets' to record electromagnetic fields and particles near the laser axis. These subset boxes spanned the full length of the $X$-axis, as well as a width of 0.2 $\mu$m in the $Y$- and $Z$-axis when dimensionality allows. The reduced data size allowed us to record these subsets every 10 fs in the simulation. We generally expect the highest energy protons to be on axis, although some features may be missed.\cite{becker2018ring}

\section{Energy Absorption and Transfer in 1D, 2D, and 3D PIC simulations} \label{sec:physical_processes}

\begin{figure*}
    \centering
    \includegraphics{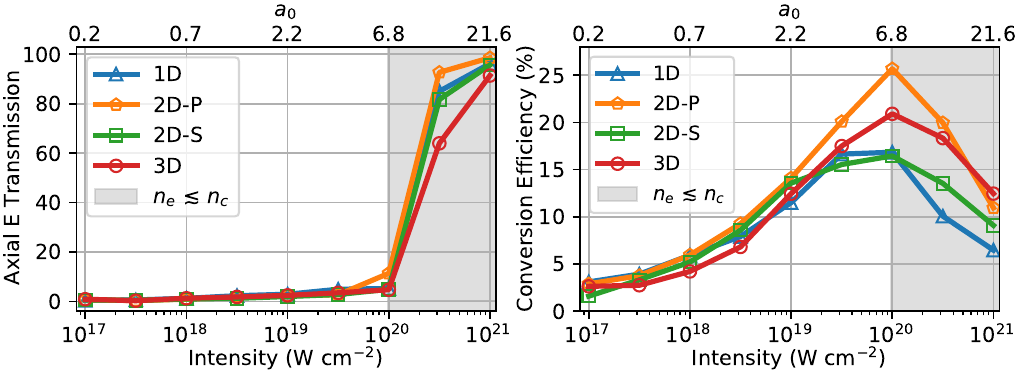}
    \caption{Relative axial electric field transmission through the target at various intensities at 130~fs after the start of the simulation is shown on the left calculated by comparing the electric field energy in the fields subset for $x>0$ (Transmission) to $x<0$ (Reflection) using the formula  (Transmission)/(Transmission + Reflection)$\times$100.  Absorption into the plasma is not considered in this calculation. The graph on the right shows total conversion efficiency from laser energy to protons and electrons  at 300~fs for each laser intensity. Conversion efficiency approximately agrees across dimensionalities for simulations with intensities of 10$^{19}$~W~cm$^{-2}$ or lower. However, once an intensity of $\sim$10$^{20}$~W~cm$^{-2}$ is reached, the total conversion efficiency begins to decrease, and disparities across dimensionality increases. [Associated dataset available at \url{https://doi.org/10.5281/zenodo.13338477}.] (Ref.~\cite{dataset_zenodo}).
    }
    \label{fig:transmission_absorption}
\end{figure*}

Figure~\ref{fig:transmission_absorption} (left) shows the transmission of light near the center of the interactions for the series of simulations. For intensities below 10$^{20}$~W~cm$^{-2}$ very little transmission is observed in all simulations. Above 10$^{20}$~W~cm$^{-2}$, relative transmission begins to increase as expected. For high intensities, the 2D-P simulations show the most relative transmission.

Figure~\ref{fig:transmission_absorption} (right) shows that the total conversion efficiency from laser energy to particle energy is similar across simulation dimensionality for lower intensities below $10^{19}$~W~cm$^{-2}$. In all cases the conversion efficiency continues to increase until the relativistic transparency threshold of $\sim 10^{20}$~W~cm$^{-2}$, where it begins to decrease. This is a similar trend to experimental observations\cite{Key_1998,Feurer_1997,Fuchs2005}, where there is an increase of laser absorption with intensity until it plateaus. Interestingly, the simulations then show a drop off of conversion efficiency, although we note that our simulations do not include a pre-plasma in front of the target as typically present in experiments. The predictions of the 2D simulations begin to diverge from each other significantly at an intensity of 10$^{19}$~W~cm$^{-2}$ with the efficiencies increasing with intensity at a greater rate for 2D-P than 2D-S and with 3D generally between the two.

\subsection{Energy Transfer}

\begin{figure*}
    \centering
    \includegraphics{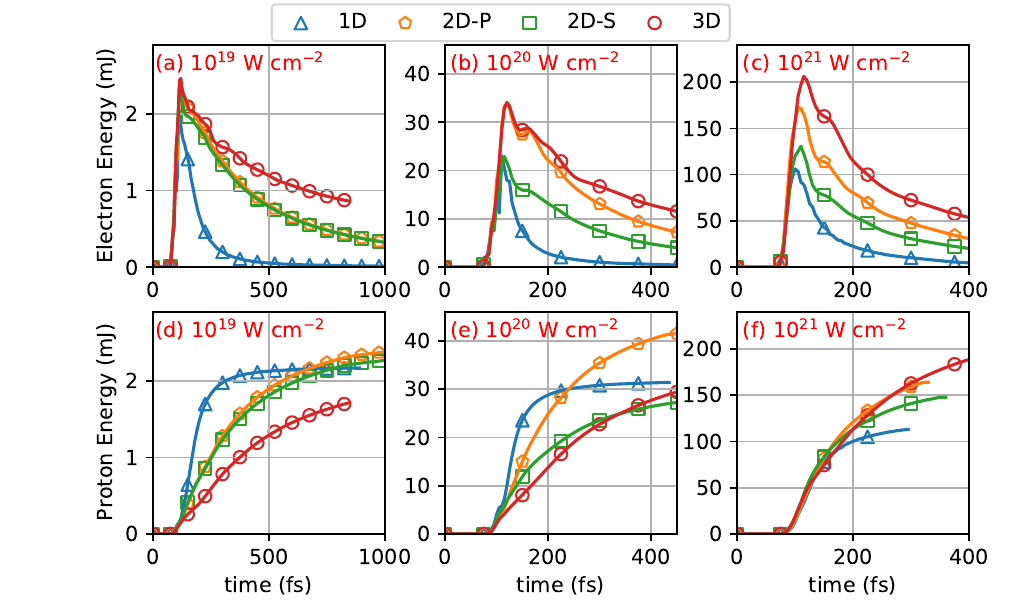}
    \caption{ Total electron energy gain vs.~simulation time is shown for intensities of  $10^{19}$~W~cm$^{-2}$ (a),  $10^{20}$~W~cm$^{-2}$ (b), and  $10^{21}$~W~cm$^{-2}$ (c). After the initial energy gain of electrons, the total electron energy decreases as energy is transferred to protons as shown in (d-f). Energy is more quickly transferred from electrons to protons in lower dimensional simulations (1D/2D) than in 3D. Graphing is stopped when energy decreases by more than 0.1\% from the previous output file. [Associated dataset available at \url{https://doi.org/10.5281/zenodo.13338477}.] (Ref.~\cite{dataset_zenodo}).  }
    \label{fig:energy_time_plots}
\end{figure*}

The laser interaction creates a charge separation between electrons and protons. Due to the differences in electric field strength over distance (Fig.~\ref{fig:field_drop}), the rate at which energy is transferred from electrons to protons will depend on simulation dimensionality. At $10^{19}$~W~cm$^{-2}$, Fig.~\ref{fig:energy_time_plots} (a) shows that the initial electron energy gain from the laser pulse is similar, but then energy is more quickly transferred from electrons to protons in 1D within a few hundred femtoseconds. In 2D and 3D, energy is still being transferred at the picosecond timescale and energy is transferred most slowly in 3D. This same behavior is exhibited at higher intensities as shown in Fig.~\ref{fig:energy_time_plots} (b-c), although differences in absorption of 2D-S and 2D-P at high intensities separates the graphs.

These differences consequently change the way that the protons gain energy due to energy conservation. For 1D simulations, energy is quickly transferred to protons and then remains fairly constant until particles begin leaving the simulation box as shown in Fig.~\ref{fig:energy_time_plots} (d-f). For 2D and 3D simulations, the energy transfer is not complete at 1~ps and the simulation run time (and box size) would need to be increased to see the proton energy plateau. This underscores that reported laser-proton conversion efficiency is strongly dependent on both the simulation run time and initial laser-electron conversion efficiency. 

For intensities up to $10^{19}$~W~cm$^{-2}$, 2D-S and 2D-P simulations show similar electron and proton energy graphs as illustrated in Fig.~\ref{fig:energy_time_plots}(a) and (d). For higher intensities, improved electron energy absorption is observed in 2D-P simulations, which results in higher total proton energy gains as shown in Fig.~\ref{fig:energy_time_plots}~(e-f).

\section{Scaling of Maximum Proton Energy}\label{sec:output_metrics}

One of the most common output metrics extracted from PIC simulations is the single highest energy of a proton macroparticle in the simulation. This maximum proton energy value is subject to uncertainty, where there are often the fewest particles in the high energy tail of the energy spectrum.\cite{Smith_2021_Compare,rehwald2023ultra} Additionally it can take many hundreds of laser periods for the maximum energy to converge, and even for large simulation boxes these highest energy particles can quickly reach the edge of the simulation box and escape before a maximum energy is obtained. 

\begin{figure*}
    \centering    \includegraphics{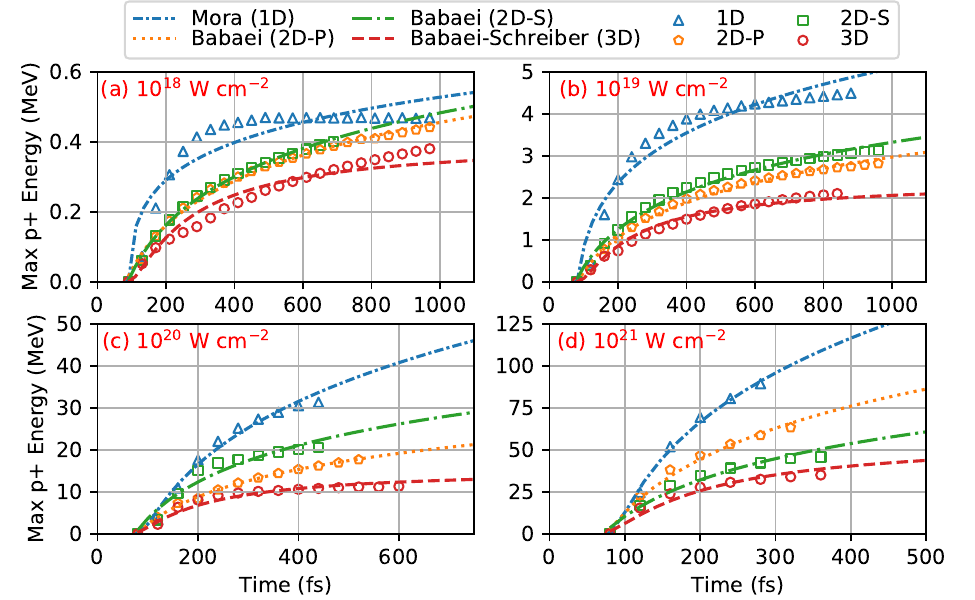}
    \caption{Hollow markers show simulation results of maximum proton energy vs.~time for each dimension with laser intensities ranging from 10$^{18}$~W~cm$^{-2}$ to 10$^{21}~$W~cm$^{-2}$. Dashed lines show the fits to Equations~\ref{3D:Babaei}-\ref{1D:mora} for the respective dimensionality. Simulation results are only shown/fit from the time when the maximum energy exceeds a threshold of 1~keV until the time when the subsequent maximum proton energy value drops by more than 3 percent. [Associated dataset available at \url{https://doi.org/10.5281/zenodo.13338477}.] (Ref.~\cite{dataset_zenodo}).    } 
    \label{fig:MaxProtonTime}
\end{figure*}

Figure~\ref{fig:MaxProtonTime} shows the maximum proton energy vs.~time for several simulations in different intensity regimes. The 1D simulations have the initial fastest growth rate of maximum ion energy, but the rate decreased after a few hundred femtoseconds. The 2D simulations show similar predictions for lower intensities, but significant discrepancies near the RIT regime. We observe different absorption and acceleration mechanisms in these regimes for the different polarizations as observed and thoroughly discussed by \citet{stark2017effects,stark_2018}.  

\subsection{Models for Maximum Proton Energy}
We will use our simulations to test the applicability of existing models of maximum proton energy over time for a wide range of simulation intensities. In 3D, \citet{Schreiber_2006_Analytical} considered a cylindrical surface charge build up and used energy conservation to find an expression for maximum proton energy. Babaei et al.\cite{babaei2017rise,sinigardi2018tnsa} then found a time dependent expression from the approach by \citet{Schreiber_2006_Analytical}, where the maximum proton energy goes as

\begin{equation}\label{3D:Babaei}
    E_{max}^{3D}(t) = E_{Fit}^{3D} \left( 1 - \frac{t^*}{t} \right)^2,
\end{equation}
for simulation times $t$ that are greater than a fitting parameter $t^*$. The maximum proton energy asymptotes to the constant fit parameter of  $E_{Fit}^{3D}$ in this model.

\citet{babaei2017rise} also extended this model to 2D, assuming a sheet of charge with finite width in the physical dimension of the simulation and infinite extent in the virtual dimension. The resulting maximum proton energy goes as
\begin{equation}\label{2D:Babaei}
    E_{max}^{2D}(t) = E_{Fit}^{2D} \ln \left(\frac{t}{t^*} \right).
\end{equation}

We perform the linearized version of these fits as described in \citet{babaei2017rise}. We also used an unmodified simulation time for $t$, which provided a good fit for our results, but the impact of this hidden parameter should be explored in the future.  For our one-dimensional simulations, we turn to the 1D plasma expansion model by Mora \cite{Mora_2003}, which can be fit with

\begin{equation}\label{1D:mora}
    E_{max}^{1D}(t) =E_{Fit}^{1D} \ln \left(\frac{t}{t^*} + \sqrt{(t/t^*)^2+1} \right)^2,
\end{equation}
where the simulation time is shifted by the time it takes for the laser pulse to reach the target. We note that the 3D model has a finite maximum proton energy, whereas the 1D, and 2D models continue to grow.

Figure~\ref{fig:MaxProtonTime} fits these three models to our simulated maximum proton energies. The 1D Mora model does not account for the saturation of maximum ion energy observed in the lower intensity simulations where particles generally stay within the simulation grid. This was also observed by \citet{djordjevic2021characterizing}. One way to address this limitation is to add an empirical acceleration stopping time as done by \citet{Fuchs2005}. The 2D Babaei model fits the data well since the 
since the 2D simulations do not equilibrate on the timescale. The 3D model provides a good fit for the relativistic intensity simulations. Clearly there are limitations of the 2D/3D models when considering some acceleration mechanisms. For example, the 3D model seems to slightly under-predict the final value at low intensities (e.g.~Fig.~\ref{fig:MaxProtonTime}(a)), and the 2D model does not capture other acceleration mechanisms such as the 2D-S simulation with an intensity of $10^{20}$~W~cm$^{-2}$  (Fig.~\ref{fig:MaxProtonTime}(c)). It is notable how well these models work over such a large order of magnitude of intensities.

\subsection{Ratios of Maximum Proton Energy}

\begin{figure}
     \centering
     \includegraphics{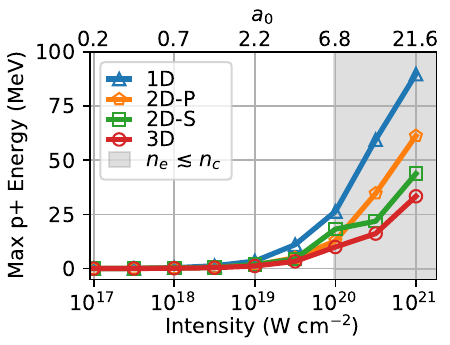}
     \caption{The maximum proton energy at 300 fs for each laser intensity and dimensionality. The 1D simulations predict the highest values followed by the 2D,  and the 3D simulations have the lowest predicted  maximum proton energies. This snapshot occurs before significant numbers of protons begin leaving the simulations. These values continue to increase with simulation time as shown in Fig.~\ref{fig:MaxProtonTime}. [Associated dataset available at \url{https://doi.org/10.5281/zenodo.13338477}.] (Ref.~\cite{dataset_zenodo}).   }
     \label{fig:MPEvsIntensity}
 \end{figure}

Figure~\ref{fig:MPEvsIntensity} shows the maximum proton energy for all simulations at 300~fs, which is a time before significant numbers of protons begin leaving the grid for the most intense simulations. Consistent with previous work\cite{sgattoni2012laser,liu2013three,d2013optimization,babaei2017rise, xiao2018multidimensional}, the 1D simulations report greater maximum proton energies than the 2D, which in turn are greater than 3D. The simulations generally have greater disagreement with larger laser intensities. 

When examining the maximum proton energy at a certain simulation time across intensities, we see greater disagreement in the RIT regime. Figure~\ref{fig:MPEvsIntensity} demonstrates that for simulations with laser intensities greater than 10$^{19}$ W cm$^{-2}$, the 1D and 2D simulations significantly overestimate the maximum proton energy. Furthermore, above 10$^{20}$ W cm$^{-2}$, the behavior of the 2D-S and 2D-P simulations diverge, with the 2D-P simulations reporting higher proton energies than the 2D-S simulations. 

\renewcommand{\arraystretch}{1.5}
\begin{table*}[ht]

\begin{tabular}{|l|ll|ll|ll|}
\hline
\multicolumn{1}{|c|}{\multirow{2}{*}{\begin{tabular}[c]{@{}c@{}}Laser Intensity\\  (W cm$^{-2}$)\end{tabular}}} & \multicolumn{2}{c|}{${E_{max}^{1D}}/{E_{max}^{3D}}$} & \multicolumn{2}{c|}{${E_{max}^{2D-S}}/{E_{max}^{3D}}$} & \multicolumn{2}{c|}{${E_{max}^{2D-P}}/{E_{max}^{3D}}$} \\ \cline{2-7} 
\multicolumn{1}{|c|}{}                                                                                          & \multicolumn{1}{l|}{t = 300 fs}    & t = 500 fs      & \multicolumn{1}{l|}{t = 300 fs}     & t = 500 fs       & \multicolumn{1}{l|}{t = 300 fs}     & t = 500 fs       \\ \hline
10$^{17}$                                                                                                       & \multicolumn{1}{l|}{2.96}          & 2.71            & \multicolumn{1}{l|}{1.05}           & 1.12             & \multicolumn{1}{l|}{1.15}           & 1.14             \\ \hline
10$^{17.5}$                                                                                                     & \multicolumn{1}{l|}{2.06}          & 1.59            & \multicolumn{1}{l|}{1.35}           & 1.29             & \multicolumn{1}{l|}{1.33}           & 1.30             \\ \hline
10$^{18}$                                                                                                       & \multicolumn{1}{l|}{2.55}          & 1.77            & \multicolumn{1}{l|}{1.54}           & 1.30             & \multicolumn{1}{l|}{1.51}           & 1.26             \\ \hline
10$^{18.5}$                                                                                                     & \multicolumn{1}{l|}{3.02}          & 2.31            & \multicolumn{1}{l|}{1.28}           & 1.24             & \multicolumn{1}{l|}{1.43}           & 1.28             \\ \hline
10$^{19}$                                                                                                       & \multicolumn{1}{l|}{2.85}          & 2.41            & \multicolumn{1}{l|}{1.54}           & 1.48             & \multicolumn{1}{l|}{1.32}           & 1.30             \\ \hline
10$^{19.5}$                                                                                                     & \multicolumn{1}{l|}{3.30}          & 3.03            & \multicolumn{1}{l|}{1.32}           & 1.33             & \multicolumn{1}{l|}{1.49}           & 1.54             \\ \hline
10$^{20}$                                                                                                       & \multicolumn{1}{l|}{2.64}          & (3.60)          & \multicolumn{1}{l|}{1.82}           & (2.15)           & \multicolumn{1}{l|}{1.28}           & 1.56             \\ \hline
10$^{20.5}$                                                                                                     & \multicolumn{1}{l|}{3.68}          & {[}(4.69){]}    & \multicolumn{1}{l|}{1.35}           & {[}(1.43){]}     & \multicolumn{1}{l|}{2.15}           & {[}(2.35){]}     \\ \hline
10$^{21}$                                                                                                       & \multicolumn{1}{l|}{2.69}          & {[}(3.29){]}    & \multicolumn{1}{l|}{1.31}           & {[}(1.39){]}     & \multicolumn{1}{l|}{1.84}           & {[}(1.97){]}     \\ \hline
\end{tabular}

\caption{Ratio of maximum proton energy for lower dimensional (1D/2D) simulations compared to 3D simulations at 300 fs and 500 fs. For the 1D and 2D simulations with greater laser intensity, the highest energy protons have left the simulation box by 500 fs (based on the criteria discussed in Fig.~\ref{fig:MaxProtonTime}) For these cases, the 1D or 2D maximum proton energy was calculated using the fit equations, Equation \ref{1D:mora} or \ref{2D:Babaei}, indicated by enclosure in parentheses (). Values in square brackets {[}{]} use Eq.~\ref{3D:Babaei} to calculate the 3D maximum proton energy when necessary.\footnote{Note: Associated dataset available at \url{https://doi.org/10.5281/zenodo.13338477}. (Ref.~\cite{dataset_zenodo}).} }\label{RatioTable}
\end{table*}

Comparing the ratios of the 1D, 2D-S, and 2D-P maximum proton energies to the 3D maximum proton energy in Table~\ref{RatioTable} provides a clearer picture of the overestimation of maximum proton energy by lower dimensional simulations. Below 10$^{18}$~W~cm$^{-2}$ laser intensity, the ratios of maximum proton energy for each of the 2D cases compared to the 3D are nearly equal and increase with laser intensity. For laser intensities ranging from 10$^{17.5}$ through 10$^{19.5}$ W cm$^{-2}$, the ratios of 2D-S to 3D and 2D-P remain relatively constant, hovering between 1.2 and 1.6 at 300~fs and 500~fs, and remain similar to each other. 

Beyond the variances in intensity, the 2D ratios also further demonstrate the discrepancy in the behavior of the two different polarizations at the high intensities as well. In lower intensity regimes, there is little difference between the maximum proton energies of the 2D-P and 2D-S simulations. This is not the case for higher intensities. At 10$^{20.5}$ W cm$^{-2}$, for example, the 2D-P maximum proton energy is over two times double that of the 3D while the 2D-S maximum proton energy is only 1.35 times higher at 300~fs. 

We can also compare the ratios of Table \ref{RatioTable} to scaling factors proposed by previous works, namely that of \citet{xiao2018multidimensional}. Xiao \textit{et al.} compared the maximum proton energy of 2D and 3D PIC simulations of TNSA and found that the ratio of these quantities could be expressed in terms of the laser beam waist $w_0$ as a constant ratio\cite{xiao2018multidimensional}
\begin{equation} \label{XiaoRatio}
    \frac{E_{max}^{2D}}{E_{max}^{3D}} = \sqrt{\frac{\pi w_0}{2}}.
\end{equation}
Using our spot size of 2 $\mu$m, Equation \ref{XiaoRatio} predicts that the ratio of 2D to 3D maximum proton energies should be 1.77. For our 2D simulations in the TNSA regime, the ratio tended to be lower than this value, with the highest ratio being only 1.54. The ratios for simulations in the RIT regime are more sporadic and do yield some higher ratios. However, several of these values also overshoot the 1.77 value calculated using Equation \ref{XiaoRatio}.

The 1D ratio appears to follow a less predictable pattern for the lower laser intensities ($\le$ 10$^{18}$ W cm$^{-2}$). The 1D to 3D ratio is consistently higher than its 2D counterparts and does not strictly increase with laser intensity. For intensities between 10$^{18}$ W cm$^{-2}$ and 10$^{19.5}$ W cm$^{-2}$, the ratio is variable, but stays within a range of 2.5 to 3.3. Then, the 1D ratio using fit values exhibits similar behavior as the 2D-P ratio, experiencing an abrupt rise followed by a sharp decrease.

\section{Conclusion}\label{sec:conclusion}
In summary, we find that key output metrics from 1D/2D PIC simulations of ion acceleration such as maximum proton energy and conversion efficiency should not be simply scaled to 3D results using a constant factor since outputs depend non-trivially on additional factors including the laser intensity. We showed that this is due both to differences in absorption of the laser pulse during the initial laser-matter interaction and to differences in energy transfer rates that are exaggerated for lower dimensional simulations. This results in lower dimensional simulations typically overestimating the maximum proton energy when compared to 3D results. Additionally, 2D-S and 2D-P simulation predictions are similar at low intensities, but diverge when nearing RIT regimes. Our results also corroborate the effectiveness and limitations of existing models of ion acceleration, which can be used to extend the predictions of a simulation beyond the end of the simulation.   

Lower dimensional simulations will continue to play an important role in the study of laser-plasma interactions. When the resources for a properly resolved 3D simulation are not available, a sufficiently high resolution 1D or 2D simulation can be used to capture important physical effects. For example, while the exact values of conversion efficiency and transmission for our simulations varied, the general trends were replicated in all dimensions. Also, the significantly reduced computational cost of 1D/2D simulations allow us to run hundreds to thousands of lower dimensional simulations in place of one three-dimensional simulation, which is especially useful for machine learning applications.\cite{smith2020optimizing, djordjevic2021characterizing,Djordjevic_2023} Additionally, there may be circumstances where a problem is sufficiently 1D or 2D in nature.\cite{Mariscal_2019}

Future work will continue to explore how dimensionality differences in different intensity regimes depend on target properties such as thickness, density profile, and geometry, as well as laser properties such as spot size, pulse duration, incidence angle, and polarization. The predictions of 1D, 2D, and 3D simulations should then be compared to the complex patterns found in experimental studies\cite{zimmer2021analysis}. Furthermore, cylindrical PIC methods are being used to model laser-wakefield and plasma-wakefield acceleration~\cite{LEHE201666} and should be compared to 2D and 3D Cartesian simulations of laser ion acceleration in the future.

\begin{acknowledgments}
This project was supported by the Appalachian Semiconductor Education and Technical (ASCENT) Ecosystem as part of the Intel Semiconductor Education and Research Program (SERP) for Ohio and the Ohio Space Grant Consortium Undergraduate STEM Scholarship. This research used resources of the National Energy Research Scientific Computing Center (NERSC), a U.S. Department of Energy Office of Science User Facility located at Lawrence Berkeley National Laboratory, operated under Contract No. DE-AC02-05CH11231 using NERSC award <FES>-ERCAP<0025066> and resources of the Ohio Supercomputer Center.\cite{OhioSupercomputerCenter1987} The code EPOCH used in this work was in part funded by the UK EPSRC grants EP/G054950/1, EP/G056803/1, EP/G055165/1 and EP/ M022463/1. This work used matplotlib~\cite{Hunter:2007}, SciPy\cite{2020SciPy-NMeth}, NumPy\cite{harris2020array}, Astropy\cite{price2022astropy}, and PlasmaPy\cite{PlamsaPy}. J.~S. would like to thank Ronak Desai for useful feedback. L.~D. would also like to thank Dennis Kuhl for useful feedback.

\end{acknowledgments}

\section*{Author Declarations}

\section*{Conflict of Interest}
The authors have no conflicts to disclose. 

\section*{Supplementary Material}
The supplementary material is a simulation input deck designed to automatically work in the 1D, 2D, and 3D versions of EPOCH, which just a change in polarization angle for the relevant 2D simulations. There are placeholder tags (e.g.,~\texttt{<intensity>})  for intensity and spot size that must be set by the user. We used an automated script to fill these in and submit simulations. Some cluster related settings such as end time were adjusted. The provided input deck includes some added and removed comments for readability.

\section*{Data Availability Statement}
The data that support the findings of this study are openly available in Zenodo at \url{https://doi.org/10.5281/zenodo.13338477}, reference number~\cite{dataset_zenodo}.

\subsection*{Author Contributions}
\textbf{Lillian A. Daneshmand}: Conceptualization (equal); Data Curation (equal); Formal Analysis (equal); Funding Acquisition (supporting); Investigation (equal); Methodology (equal); Project Administration (supporting); Validation (equal); Visualization (equal); Writing/Original Draft Preparation (equal); Writing/Review \& Editing (equal). \textbf{Madeline Aszalos}: Formal Analysis (supporting); Methodology (equal); Visualization (equal); Writing/Original Draft Preparation (supporting); Writing/Review \& Editing (supporting). \textbf{Scott Feister}: Project Administration (supporting); Resources (equal); Supervision (supporting); Visualization (equal); Writing/Original Draft Preparation (supporting); Writing/Review \& Editing (supporting). \textbf{Joseph R. Smith}: Conceptualization (equal); Data Curation (equal); Formal Analysis (equal); Funding Acquisition (lead); Investigation (equal); Methodology (equal); Project Administration (lead); Resources (equal); Supervision (lead); Validation (equal); Visualization (equal); Writing/Original Draft Preparation (equal); Writing/Review \& Editing (equal).

\appendix

\section{Considerations for Laser Focusing and Energy Scaling}\label{app:focus_energy}

\begin{figure}
    \centering
    \includegraphics{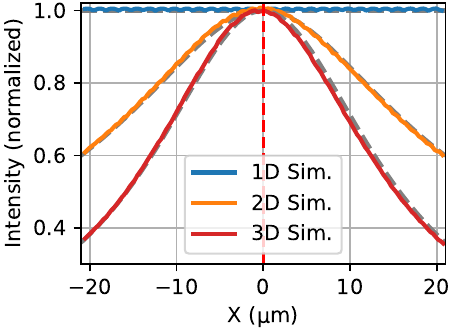}
    \caption{Normalized axial intensity of a Gaussian laser pulse coming to focus at $x=0$ for 1D, 2D, and 3D PIC simulations in vacuum.  Solid lines are from simulations and dashed lines are Equations~\ref{eq:focus_3d}-\ref{eq:focus_1d}. [Associated dataset available at \url{https://doi.org/10.5281/zenodo.13338477}.] (Ref.~\cite{dataset_zenodo}).  }
    \label{fig:focus_las}
\end{figure}

For lower dimensional PIC simulations, components of fields in the `virtual dimensions' are still modeled. Additionally the particle momenta are updated for each physical and virtual dimension. As such these codes are often referred to as 1D(3v), 2D(3v), and 3D(3v), for brevity we refer to these as 1D, 2D, and 3D in this work. The propagation of electromagnetic waves depends on simulation dimensionality, where the axial intensity of a Gaussian beam propagating along $x$, goes as
\begin{align}\label{eq:focus_3d}
    I_{3D}(X) &= I_0/\left(1+ \left(\frac{X}{X_r}\right)^2\right),\\ 
    I_{2D}(X) &= I_0/\sqrt{1+ \left(\frac{X}{X_r}\right)^2},\\
  \label{eq:focus_1d}   I_{1D}(X) &= I_0,
\end{align}

\noindent where $I_0$ is the intensity at focus and $X_r = \pi w_0^2/\lambda$ is the Rayleigh length with $w_0$ being the beam waist radius, and $\lambda$ is the laser wavelength\cite{Ngirmang_2016}. Lasers do not focus in 1D PIC simulations, and focus more weakly in 2D than 3D\cite{Ngirmang_2016}. For simulations with a thin target placed at the focus of the pulse, the on-target intensity will be similar in all three cases. We expect more significant implications when the interaction region is not confined to the focal spot including cases of extended pre-plasma \cite{Ngirmang_2016,gu2021multi}, the use of structured targets~\cite{Jiang_Structured_2016,blanco2017table,wang_blackman_2021}, and when the laser is transparent to the target~\cite{stark2017effects}. Figure~\ref{fig:focus_las} illustrates this difference. The solid lines on the graph are from PIC simulations (as described in section~\ref{sec:sim_setup}, but with the target removed) by tracking a peak in the electric field near the center of the laser pulse.

The total energy in a Gaussian laser pulse in a PIC simulation also depends on the simulation dimensionality ($N = 1, 2, 3)$ as 

\begin{equation}
    E_{Laser,ND} = \left(w_0 \sqrt{\pi /2}\right)^{N-1}  I_0 \times  F(\tau_{laser}) \times (L_V)^{3-N},
\end{equation}
where for a sine-squared temporal profile $F(\tau_{laser}) = \tau_{FWHM}$, and the length of the virtual dimension(s) may be $L_V$=1~m, but depends on how units are treated in the PIC code~\cite{Smith_2021_Compare}. 

The total initial particle kinetic energy in a PIC simulation can be calculated by multiplying the average particle kinetic energy (e.g.~$3/2\times k_B T$, where $k_B$ is the Boltzmann constant and $T$ is the initial target temperature) by the total number of particles in the simulation (for each species). To calculate the number of particles in 1D/2D, the extent of the virtual dimension must be included in the calculation for number of particles.

For this work we simply take the initial particle energy reported by the simulation, as small variations come from sampling a finite number of particles from the Maxwell-Boltzmann distribution. When calculating conversion efficiencies and energy gain, the initial particle energy is subtracted away from the particle energy at a later time step. This is an important correction for low intensities and high initial particle temperatures. Similarly, the total laser energy on grid can be found numerically if the simulation box is large enough, or the injected energy is tracked, as there can be minor differences between expected and numerical values.

\bibliography{0_Dimension}

\end{document}